\documentclass[10pt,preprint]{aastex}

\usepackage{graphicx,natbib}

\shorttitle{Karhunen-Lo\`eve Image Projection (KLIP)}
\shortauthors{Soummer, Pueyo \& Larkin}

\begin{document}

\title{Detection and Characterization of Exoplanets and Disks\\ using Projections on Karhunen-Lo\`eve Eigenimages}

\slugcomment{Accepted for publication in ApJ Letters July 16, 2012}

\author{R\'emi Soummer}
\affil{Space Telescope Science Institute, 3700 San Martin Drive, Baltimore MD 21218, USA}
\email{email: soummer@stsci.edu}

\author{Laurent Pueyo\altaffilmark{1}}
\affil{Department of Physics and Astronomy, Johns Hopkins University, Baltimore, MD, USA}
\email{lap@phg.jhu.edu}

\author{James Larkin}
\affil{Physics and Astronomy, University of California Los Angeles, CA, USA}
\email{larkin@astro.ucla.edu}

\altaffiltext{1}{Space Telescope Science Institute, 3700 San Martin Drive, Baltimore MD 21218, USA}

\begin{abstract}
We describe a new method to achieve point spread function (PSF) subtractions for high-contrast imaging using Principal Component Analysis (PCA) that is applicable to both point sources or extended objects (disks). 
Assuming a library of reference PSFs, a Karhunen-Lo\`eve transform of theses references is used to create an orthogonal basis of eigenimages, on which the science target is  projected. 
For detection this approach provides comparable suppression to the Locally Optimized Combination of Images (LOCI) algorithm, albeit with  increased robustness to the algorithm parameters and speed enhancement. 
For characterization of detected sources the method enables forward modeling of astrophysical sources. This alleviates the biases in the astrometry and photometry of discovered faint sources, which are usually associated with LOCI-based PSF subtractions schemes. We illustrate the algorithm performance using archival Hubble Space Telescope (HST) images, but the approach may also be considered for ground-based data acquired with Angular Differential Imaging (ADI) or  integral-field spectrographs (IFS).
\end{abstract}


\maketitle 
\section{Introduction}\label{sect:intro}

Imaging faint exoplanets and circumstellar disks is a difficult task limited by the small angular separation and high contrast between the faint astrophysical signal and the residual starlight \citep{OH09}. The first direct images of planets or other interesting localized structures orbiting nearby stars were obtained relatively recently \citep{MMB08,KGC08,LGC09,LBC10,MZK10,Kalas11,Janson12,KI12}. 
Direct imaging results will significantly expand in the near future with a new generation ground-based instruments being implemented \citep{MGP08,BFD08,HOZ11,HST08}.
Even using advanced starlight suppression systems (e.g. coronagraphs) such as the one designed for the Gemini Planet Imager \citep{SKM01,S05,SSP11}, image post-processing remains necessary to achieve the full high-contrast performance \citep{MMS08}. 
Indeed, most recent direct imaging results were enabled by key advances in observational procedures, e.g. Angular Differential Imaging (ADI) \citep{MLD06}, simultaneous imaging of contiguous wavelengths \citep{MDR00,SF02,PCV12},  in conjunction with sophisticated images analysis routines based on the Locally Optimized Combination of Images (LOCI) algorithm \citep{LMD07}. 
Although LOCI is extremely powerful at detecting point sources, the very aggressive PSF subtraction introduces biases on the photometry and astrometry, which have to be calibrated carefully \citep{MMV10}. 
Biases can be reduced significantly using statistical exploration of the algorithm parameter space \citep{SHP11}, a computationally expensive solution, or by modifying the underlying cost function by masking some pixels \citep{MMV10} or using damping penalty terms \citep{PCV12}. 
LOCI tends to subtract extended structures such as debris disks, and several studies have focused on mitigation strategies, taking advantage of the geometry in edge-on disks or performing an optimization over large zones  \citep{FKD07,TGG10,BTV10}. 

In this paper we introduce a new reduction algorithm based on projections on a truncated Karhunen Lo\`eve  transformation of the reference PSF library. We first introduce the general formalism associated with PSF subtraction and describe the algorithm. We then  use HST NICMOS coronagraphic data to illustrate its performances and highlight its advantages for the characterization of point sources and the discovery of faint extended structures around nearby stars.

\section{The KL Image Projection (KLIP) algorithm}

\subsection{Optimal PSF subtraction}
An astronomical Point Spread Function (PSF) can be seen as a $N$-dimensional stochastic process indexed over a scalar quantity $\psi =\psi(t,\lambda,\theta,\phi,T_{exp},m_{Star})$, which is representative of  the state of the instrument-telescope system (for instance: time, wavelength, field rotation angle, wavefront error, exposure time, star magnitude) at the time of the exposure. With $N$ pixels in the image, the PSF intensity can be written as: 
\begin{equation}
 I_{\psi}(n)=I(p[n],q[n],\psi) , \label{eq1}
\end{equation}
where we have rearranged the detector coordinates in one dimension using the pixel mapping functions $p$ and $q$ and the pixel index $n$ $(n \in [1,n])$. 

We assume an observation sequence that yields a target image: 
\begin{equation}
T(n) = I_{\psi_0}(n) + \epsilon \, A(n),
\end{equation}
which might ($\epsilon =1$) or might not ($\epsilon =0$) contain faint astronomical signal $A(n)$, 
and a set of reference images: 
\begin{equation}
\{ R_k(n) =   I_{\psi_k}(n) \}_{k = 1...K}. 
\end{equation}
$\psi_{0}(n)$ and $\psi_{k}(n)$ represent the realizations of the state of the telescope-instrument system respectively for the target image and the set of references. 

Using prior information about the observation strategy, these references have been selected so that they do not contain any astronomical signal in the search region $\mathcal{S}$ (dimension $N_{\mathcal{S}}$ pixels) of the PSF. Moreover the observations have been carried out at neighboring states of the telescope/instrument system: $|\psi_{i}-\psi_{j}| <\delta$ for all $i,j \in [0,K]$, where $\delta$ is a small quantity that can be adjusted in practice using the correlation between images as a proxy. 

The goal of PSF subtraction is to reconstruct $I_{\psi_0}(n)$ using the set of reference images and to subtract it from the target image since $\epsilon A(n) = T - I_{\psi_0}(n)$. 
Unfortunately  $ I_{\psi}(n)$ is a continuous random process of $\psi$ and thus an infinite amount of references would be needed to reconstruct $I_{\psi_0}(n)$ exactly.
In practice we can only seek to determine the best estimate $\hat{I}_{\psi_0}(n)$ of $I_{\psi_0}(n)$ using  the limited information contained in the K references. 

Assuming that each reference is unique, the ensemble $\{ I_{\psi_k} \}_{k = 1...K}$ spans $\mathcal{I}_{K}$, a K dimensional sub-space of $\mathbb{R}^{N_{\mathcal{S}}}$, which has an infinity of orthonormal bases $\{ Z_k \}_{k = 1...K}$. This problem of estimating $\hat{I}_{\psi_0}(n)$ can be formulated as:\\ 

{\em What is the basis set $\{ Z_k(n) \}_{k = 1...K}$ most likely to minimize the distance between a random realization of $I_{\psi}(n)$ ($\psi = \psi_0$ in our case) and  $\mathcal{I}_{K}$? More formally:}
\begin{equation}
 \min_{\{Z_k\}} \left\{  E_{\psi}\left[ \sum_{n = 1}^{N_{\mathcal{S}}} \left(I_{\psi} - \sum_{k = 1}^{K} <I_{\psi},Z_k>_{\mathcal{S}}  Z_k (n)\right)^2 \right] \right\}\label{eq2}
\end{equation}
 where $ E_{\psi}[\cdot]$ stands for the expected value over the telescope realizations and $<\cdot,\cdot>_{\mathcal{S}} $ denotes the inner product over $\mathcal{S}$. This is a well known problem in signal processing and, assuming that all the images are of zero mean over $\mathcal{S}$, the optimal basis set $\{ Z^{KL}_k\}_{k = 1...K}$ is given by the Karhunen-Lo\`eve transform of the $\{ I_{\psi_k} \}_{k = 1...K}$ \citep{Karhunen,Loeve}. Moreover if one chooses to only describe $I_{\psi_0}$ over a $K_{klip} <K$ dimensional sub-space, then the optimal basis set is given by the truncated Karhunen-Lo\`eve transform of the references, $\{ Z^{KL}_k\}_{k = 1...K_{klip}}$\\
 \subsection{The algorithm}
The KLIP algorithm consists of the following steps:
\begin{enumerate}
\item partition the target $T(n)$ and references $R_k(n)$ images in an ensemble of search areas $\mathcal{S}$, and subtract their average values so that they have zero mean over each $\mathcal{S}$.
\item compute the Karhunen-Lo\`eve transform of the set of reference PSFs $\{R_k(n) \}_{k = 1...K}$:
\begin{equation}
 Z^{KL}_k(n) = \frac{1}{\sqrt{\Lambda_k}}\sum_{p=1}^{K} c_k(\psi_p) R_p(n) ,\label{eq6}
\end{equation}
where the vectors $C_k = [c_{k}(\psi_1) ... c_{k}(\psi_K)]$ are the eigenvectors of the K-dimensional covariance matrix $E_{RR}$ of the $R_k(n)$ references over $\mathcal{S}$, and $\{\Lambda\}_{k=1...K}$ are its eigenvalues.
\begin{eqnarray}
 E_{RR}[i,j] &=& \sum_{n = 1}^{N_{\mathcal{S}}} R_i(n)R_j(n) \label{eq4}\\
 E_{RR}\cdot C_k &=& \Lambda_k C_k, \; \Lambda_1>\Lambda_2>....>\Lambda_N \label{eq5}
 \end{eqnarray}

\item choose a number of modes $K_{klip}$ to keep in the estimate $\hat{I}_{\psi_0}(n)$.  
\item compute the best estimate  $\hat{I}_{\psi_0}(n) $ of the actual PSF $I_{\psi_0}(n) $ from the projection of the science target $T(n)$ on the  $\{Z_k^{KL} (n)\}_{k = 1...K_{klip}}$ Karhunen-Lo\`eve basis:
\begin{equation}
\hat{I}_{\psi_0}(n) =\sum_{k = 1}^{K_{klip}} <T,Z^{KL}_k>_{\mathcal{S}} Z^{KL}_k(n).
\end{equation}

 \item calculate the final image $   F(n) = T(n) - \hat{I}_{\psi_0}(n)$,
 \begin{eqnarray}
  F(n) &=& \left(I_{\psi_0}(n) - \sum_{k = 1}^{K_{klip}} <I_{\psi_0},Z^{KL}_k>_{\mathcal{S}}  Z^{KL}_k (n)\right) \nonumber\\
  &+& \epsilon \left(A(n) - \sum_{k = 1}^{K_{klip}} <A,Z^{KL}_k>_{\mathcal{S}}  Z^{KL}_k (n)\right). \label{eq::RandS}
 \end{eqnarray}
\end{enumerate}

\subsection{Throughput and Signal to Noise Ratio}\label{Sec:SNR}
\begin{figure}[htbp]
\center
\resizebox{0.7\hsize}{!}{\includegraphics{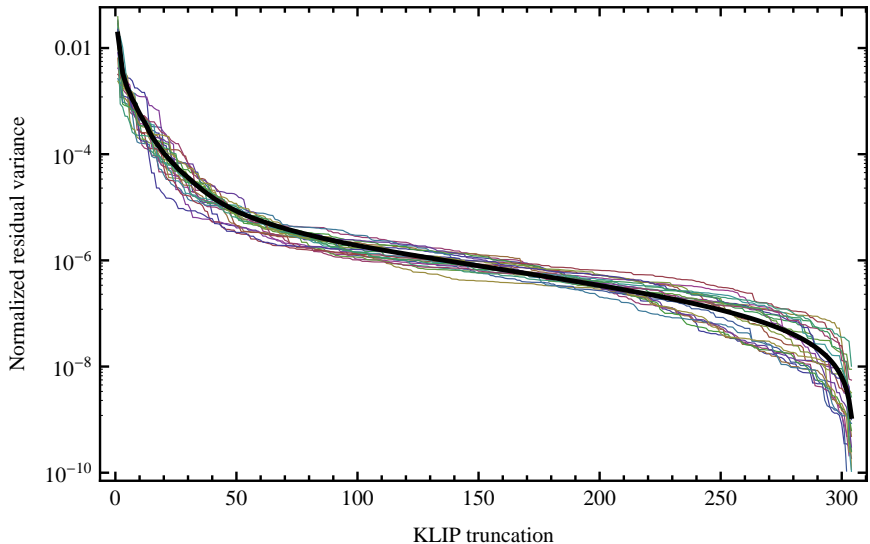}}
 \caption{Normalized total residual variance over the entire image after PSF subtraction. The black curve corresponds to the theoretical expression $\sigma^2_S/||I_{\psi_0}||^2$ based on the corollary of the  Karhunen-Lo\`eve theorem (See Equation \ref{Eq:variance}). The color curves correspond to the empirical total residual variance after subtraction for several target stars from the HST NICMOS archive without any astrophysical signal around them. In the case of a target with an astrophysical signal, the variance saturates to a minimum level and does not follow the theoretical expression.  The theoretical expression from Equation \ref{Eq:variance} can therefore be used to estimate semi-analytically the residual noise after subtraction. If KLIP is applied in small regions (e.g. using zones as with LOCI) this expression can be used to estimate the noise locally around a putative source in the target image.}\label{Fig:variance}
\end{figure}

Equation \ref{eq::RandS} provides fundamental information pertaining to the optimal choice of $K_{klip}$ for a given search zone $\mathcal{S}$. The variance in the search zone $\mathcal{S}$ after the PSF subtraction is given by: 
\begin{eqnarray}
\sigma^2_\mathcal{S} &=& ||I_{\psi_0}||^2 \left({1 -  \sum_{k = 1}^{K_{klip}}\left( \frac{ <I_{\psi_0},Z^{KL}_k>_{\mathcal{S}}}{||I_{\psi_0}||} \right)^2}\right)\\
& \simeq & ||I_{\psi_0}||^2 \left({1 -  \frac{\sum_{k = 1}^{K_{klip}} \Lambda_k}{\sum_{k = 1}^{K} \Lambda_k}}\right),\label{Eq:variance}
\end
{eqnarray}
which is a decreasing function of $K_{klip}$. 
In Figure \ref{Fig:variance} we show the impact of $K_{klip}$ on the residual variance integrated over entire KLIP-reduced images (i.e. $\mathcal{S}$ is the full image), using  HST-NICMOS coronagraphic data preselected to have no astrophysical signal. Figure \ref{Fig:variance} shows that the empirical residual variance after PSF subtraction follows the theoretical estimate in Equation~\ref{Eq:variance}.

\begin{figure}[htbp]
\center
\resizebox{0.7\hsize}{!}{\includegraphics{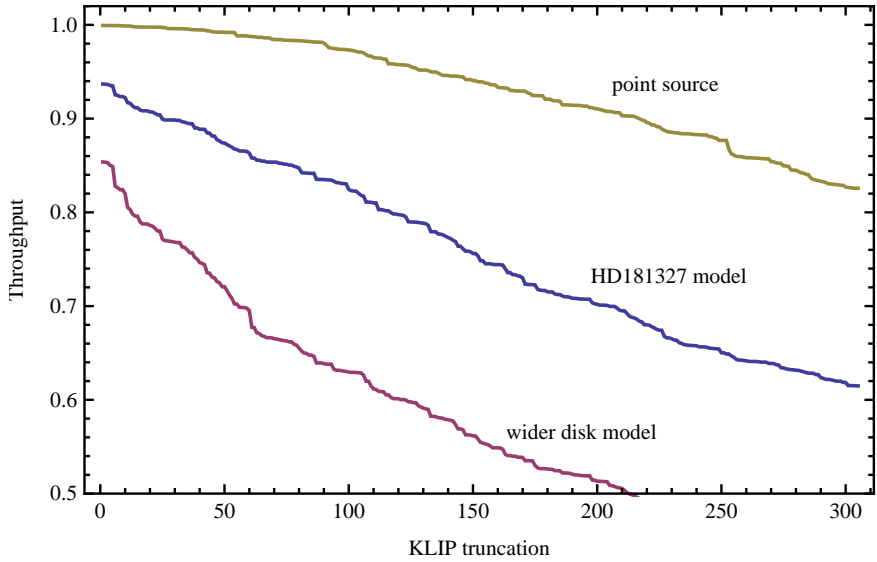}}
 \caption{KLIP throughput based on the expression in Equation~\ref{eq::RandS} from the projection of the astrophysical signal $A$ onto the KL basis $<A,Z^{KL}_k>_{\mathcal{S}}$. Since a planet's signal is mostly orthogonal to the main KL modes the throughput is high, as shown for a planet located at $\sim1.5$ arcsec from the star (top curve). For more extended sources the throughput is reduced, as illustrated for a simple geometrical model corresponding to the HD 181327 disk (middle curve). The bottom curve corresponds to a hypothetical ring-like disk about three times wider than  HD 181327. Note that here $\mathcal{S}$ corresponds to an entire HST NICMOS image, the throughput will be smaller when using smaller partitions of the image. However because it is calculated analytically for a source model independent from its flux, it can be calibrated easily.}\label{Fig:throughput}
\end{figure}

Similarly, equation \ref{eq::RandS} provides a simple expression for the signal: 
\begin{equation} 
S =  \sum_{ A(n) \neq0} \left( A(n) -  \sum_{k = 1}^{K_{klip}} <A,Z^{KL}_k>_{\mathcal{S}} Z^{KL}_k(n) \right) \label{Eq:signal}
\end{equation}
For the dominant  modes (i.e. $k \ll K$) $<~A,Z^{KL}_k>_{\mathcal{S}}\,\simeq  0$ since the morphology of the astronomical signal cannot be reproduced by a PSF realization. 
Indeed the PSF of a faint planet is localized over a few pixels and thus almost orthogonal to the main modes of the telescope's PSF realizations. Naturally this argument depends on the spatial morphology of the signal: in Figure~\ref{Fig:throughput} we show that the algorithm throughput decreases when $K_{klip}$ increases for several types of faint astrophysical object. 
Given priors on the shape of the astrophysical signal (i.e. planet position and flux, or a disk model), it is possible to estimate directly the Signal to Noise Ratio (SNR) as a function $K_{klip}$ when $\mathcal{S}$ is small enough so that  equation \ref{Eq:variance} represents the local noise in the vicinity of the source. It is therefore possible to choose $K_{klip}$ {\em a priori} in order to optimize the SNR for a given putative source in the image. 

\section{Discussion}

\subsection{Forward Modeling with KLIP}

In practice the optimal search area $\mathcal{S}$ might be too large to use Equation~\ref{Eq:variance} and Equation~\ref{Eq:signal} to pre-select $K_{klip}$. Instead $K_{klip}$ can be tuned by inspecting the SNR of putative sources in reduced images. 
However the second term of Equation~\ref{eq::RandS} carries all the information necessary to {\em characterize} faint astrophysical signal once it has been detected. Indeed the impact of the algorithm on the signal only depends on the projection of $A(n)$ on the $\{Z^{KL}_k\}_{k=1..K_{klip}}$ and is not a function of $I_{\psi_0}$. As a consequence KLIP enables forward modeling of detected planets and disks by fitting directly an astrophysical model $A_{\xi}(n)$ to the final reduced images $F(n)$, where $\xi$ represents the model parameters (e.g. astrometry and photometry for a planet, or geometry, grain distribution for a disk). $\xi$ can be obtained by minimizing the following cost function: 
\begin{equation}
 \min_{\{\xi\}} \left\{  \sum_{n = 1}^{N_{\mathcal{S}}} \left(F(n)-A_{\xi}(n) + \sum_{k = 1}^{K_{klip}} <A_{\xi},Z^{KL}_k>_{\mathcal{S}}  Z^{KL}_k(n)\right)^2\right\}.\label{Eq:fowardMod}
\end{equation}
A similar approach was proposed by \citet{MMV10} using LOCI, but it presents degeneracies because the LOCI coefficients are not guaranteed to be independent of $I_{\psi_0}$ or $A$.  Such degeneracies do not arise in the case of  KLIP because the effect of the algorithm on the astrophysical source is completely determined from Equation~\ref{Eq:fowardMod} and because the basis  $\{Z_k^{KL} (n)\}_{k = 1...K_{klip}}$ is independent of the science target. Note that Equation~\ref{Eq:fowardMod} assumes a noiseless signal. Optimal observers \citep{KB06,CBD07,CBR09} can mitigate the impact of noise in the final estimate of $\xi$, and their implementation is greatly facilitated by the fact that Equation~\ref{Eq:fowardMod} only consists of propagating $A_{\xi}(n)$ through a linear filter. Discussion regarding the practical implementation and performances of such observers are however beyond the scope of this study. 

\begin{figure}[htbp]
\center
\resizebox{0.7\hsize}{!}{\includegraphics{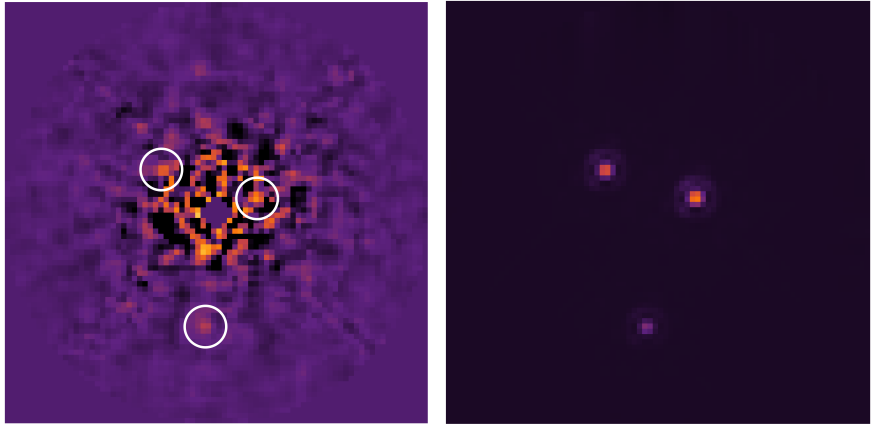}}
 \caption{Left: Final KLIP-subtracted image from HR8799 using HST NICMOS archival data from 1998. The result is obtained by applying KLIP locally in large annular zones using the first 42 KL images. Right: result from forward modeling using Equation \ref{Eq:fowardMod}. Note that the left image corresponds to a single reduction and is comparable in contrast to single images obtained previously with LOCI \citep{SHP11}. Deeper images can be obtained by combining data from different orients and by exploring the algorithm parameter space. }\label{Fig:HR8799}
\end{figure}

The left panel of  Figure~\ref{Fig:HR8799} shows an archival HST NICMOS image of HR 8799 reduced with KLIP using annular search zones. While the three planets $b$,$c$, and $d$ are visible they are not detected unambiguously: the result reported by \citet{SHP11} used a combination of  thousands of LOCI reductions, and took advantage of two HST roll angles to minimize the false alarm probability. One could envision a similar parameter search with KLIP. 
The computational cost of such an exercise would however be reduced since KLIP does not require the introduction of ``optimization zones" and ``subtraction zones" \citep{LMD07}, which reduces the dimensionality of the algorithm's parameter exploration. 
More importantly the astrometry reported in \citet{SHP11} was the result of a lengthy characterization of the biases associated with LOCI, involving thousands of reductions with synthetic companions. On the right panel of Figure~\ref{Fig:HR8799} we illustrated how one can determine the astrometric location and the photometry of HR8799bcd with forward modeling in a single KLIP-reduced image. 
In this example we performed a simple $\chi^2$ minimization of the position and flux of each planets, but this framework enables easy implementation of more appropriate modeling using Markov Chains Monte Carlo (MCMC). 
KLIP thus provides astrophysical estimates which do not contain any bias arising from self-subtraction in the reduction algorithm. The uncertainties on the astrophysical parameters are solely driven by the local noise in the reduced image. This can be addressed by optimizing the KLIP parameters to increase local SNR, using several exposures and taking advantage of observers that are optimal with respect to the residual noise. 

 \subsection{Similarities between LOCI and KLIP}
PCA has been extensively used in astronomy to reduce the dimensionality of large datasets, e.g. \citet{CAM09,CSB95}. Principal components can also be used directly with LOCI or any of its variants to reduce the dimensionality of the reference set, using KL eigen-images instead of reference PSFs. 

The cost function associated with KLIP (Equation~\ref{eq2}) is very similar to the cost function associated with LOCI \citep{LMD07}. 
However, LOCI minimizes the distance between the science target and the reference PSFs, whereas KLIP minimizes, in the statistical sense, the distance between {\em any} PSFs $I_{\psi}$ in the close vicinity of  references library and the $K_{klip}$-dimensional orthonormal basis-set $\{Z_k^{KL} (n)\}_{k = 1...K_{klip}}$. 
 It is in principle possible to include the target to the reference library before decomposition into principal components, but in this case astrophysical sources can be completely subtracted and forward modeling cannot be carried out. While  such solutions might yield slightly deeper contrasts they might yield astrophysical estimates with biases that are difficult to calibrate. In this sense LOCI, with well chosen parameters, might be optimal for detection while KLIP is more robust for characterization.

Because the set of $\{R_{k}\}_{k=1...N}$ is not an orthonormal basis, LOCI requires the inversion of the covariance matrix $ E_{RR}$, which is typically ill-conditioned. 
This leads to noise amplification if the inversion is not properly regularized. Several approaches to regularization exist, for example eigenvalue truncation \citep{MMV10}, or Tikhonov regularization \citep{PCV12,SHP11}. 
With KLIP the regularization is implicit with the truncation of the KL basis. It is also more easily controlled because the KL set remains optimal for any value of $K_{klip}$.
In addition, simple regularization schemes do not reduce the dimensionality of the covariance matrix, whereas KLIP reduces the number of references and therefore improves speed. 
Once the KL transform has been generated for a given search area geometry, the remaining free parameter $K_{klip}$ can be explored with minimal computation burden (projections).
Depending on the implementation of the algorithm and the type of data, KLIP can lead to considerable speed increases, especially when considering KLIP does not require to process synthetic sources for characterization of detected signal. 

\begin{figure}[htbp]
\center
\resizebox{0.7\hsize}{!}{\includegraphics{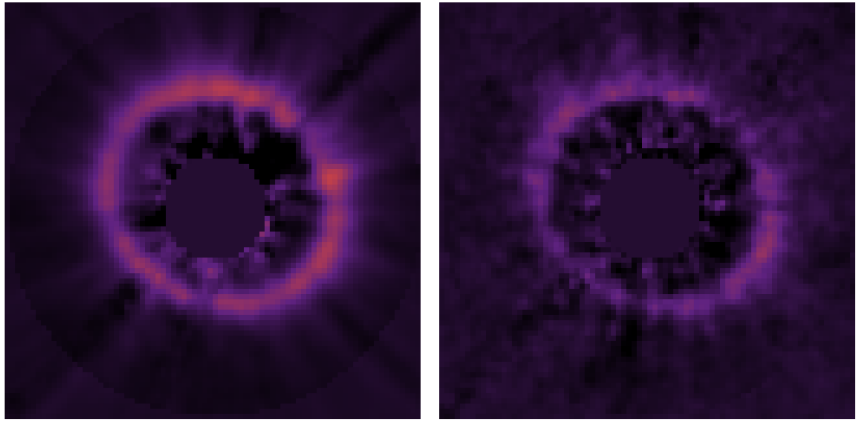}}
 \caption{
 left: image of HD 181327 using KLIP from a carefully selected reference library of 232 PSFs, using 35 KLIP coefficients. For comparison the right panel shows the image obtained with LOCI without regularization using the same dataset. We were able to obtain a very similar image using appropriate regularization with LOCI. However, there is no guarantee that the structure detected in a reduced LOCI image is not an artifact of the regularization. KLIP circumvents this degeneracy and enables forward modeling. }\label{Fig:DisksSims}
\end{figure}

As far as detection is concerned the most striking difference between LOCI and KLIP resides in disk imaging. Using HST Cycle 10 NICMOS archival data of the disk around HD  181327, we performed a global-LOCI and global-KLIP reduction over the entire image, using a carefully pre-selected reference library of 232 PSFs. The disk edges imaged with KLIP, Fig.~\ref{Fig:DisksSims}, are significantly better constrained when compared with state of the art classical PSF subtraction \citep{SSB99}. For comparison we show in Figure \ref{Fig:DisksSims} a LOCI reduction of the same object without any regularization: the self subtraction degrades the rich astrophysical information contained in the ring's edge.  It is important to note that with appropriate regularization it is possible to obtain with LOCI a very similar result to KLIP. However, with LOCI it is not easy to model the astrophysical signal since there is no guarantee that the algorithm configuration optimal for detection will not bias the astrophysical signal. 

\section{Conclusion}
In this paper we have introduced a new algorithm for high-contrast image post-processing. KLIP is a cousin of existing methods (e.g LOCI) in the sense that it builds, using a quadratic cost function, a synthetic reference PSF as a linear combination of references from a library. KLIP involves a truncated Karhunen-Lo\`eve transform of the reference library in order to produce an optimal set of orthogonal references. The science target is then projected onto a truncated set of these eigenimages to generate the synthetic reference. We used HST NICMOS data to show that KLIP provides detectability performances similar to LOCI. 
In addition to reducing the dimensionality of the data with the KL truncation, KLIP provides a linear framework that enables forward modeling of astrophysical sources by fitting directly an astrophysical model to the PSF-subtracted data, without introducing degeneracies. For example the algorithm throughput for a planet candidate can be calculated semi-analytically and independently of the planet signal without having to inject synthetic companions in a reference PSF. 
Taking advantage of the properties of the KL transform, it is possible to generate a theoretical expression of the SNR in subtracted images, provided that KLIP is applied is sufficiently small zones to be representative of the local noise around the planet candidate. 
 The algorithm was illustrated using HST data, but it is also applicable to ground-based IFS data, and preliminary tests show that a local application is preferable in this case because of the lesser degree of correlation between PSFs. 


\acknowledgements  
The authors also thank J. B. Hagan, M. Perrin, S. Lonsdale, T. Comeau and M. Russ for work and help on pipeline development;  G. Schneider and D. Hines for discussions on disk imaging with NICMOS, and for providing the LAPL reference PSF library. The authors also thank C. Chen, M. Fitzgerald, B. M\'enard,  C. Norman, B. Oppenheimer, A. Sivaramakrishnan, A. Szalay, G. Vasisht for discussions. 
Support for Program number HST-AR-12652.01-A was provided by NASA through a grant from the Space Telescope Science Institute, which is operated by the Association of Universities for Research in Astronomy, Incorporated, under NASA contract NAS5-26555. This work was also performed in part under contract with the Jet Propulsion Laboratory (JPL) funded by NASA through the Sagan Fellowship Program. JPL is managed for NASA by the California Institute of Technology.


\end{document}